\documentclass[12pt,onecolumn,preprintnumbers,amsmath,amssymb]{revtex4}

\usepackage{graphicx}
\usepackage{dcolumn}
\usepackage{bm}
\usepackage{color}

\begin{document}


\title{The effects of fluid viscosity on the kinematics and material properties of \emph{C. elegans} swimming at low Reynolds number}

\author{J. Sznitman$^{1,2}$}
\author{X. Shen$^1$}%
\author{Prashant K. Purohit$^1$}%
\author{P.E. Arratia$^1$}\email{parratia@seas.upenn.edu}
\affiliation{%
\small $^1$Department of Mechanical Engineering \& Applied Mechanics, University of Pennsylvania, Philadelphia PA 19104, USA\\
\small $^2$Present Address: Department of Mechanical \& Aerospace Engineering, Princeton University, Princeton NJ 08544, USA}


\date{\today}

\begin{abstract}
\textbf{Abstract:} The effects of fluid viscosity on the kinematics of a small swimmer at low Reynolds numbers are investigated in both experiments and in a simple model. The swimmer is the nematode \emph{Caenorhabditis elegans}, which is an undulating roundworm approximately 1~mm long. Experiments show that the nematode maintains a highly periodic swimming behavior as the fluid viscosity is varied from 1.0~mPa$\cdot$s to 12~mPa$\cdot$s. Surprisingly, the nematode's swimming speed ($\sim 0.35$~mm/s) is nearly insensitive to the range of fluid viscosities investigated here. However, the nematode's beating frequency decreases to an asymptotic value ($\sim 1.7$~Hz) with increasing fluid viscosity. A simple model is used to estimate the nematode's Young's modulus and tissue viscosity. Both material properties increase with increasing fluid viscosity. It is proposed that the increase in Young's modulus may be associated with muscle contraction in response to larger mechanical loading while the increase in effective tissue viscosity may be associated with the energy necessary to overcome increased fluid drag forces.

\end{abstract}

\pacs{Valid PACS appear here}
\maketitle

\section{Introduction}

Many, if not most, living organisms move in the realm of low Reynolds ($Re$) number, typically defined as $Re=\rho U L/\mu$, where $\rho$ and $\mu$ are the fluid density and viscosity respectively, and $U$ and $L$ are the organism's characteristic speed and length scale. Due to their small size, microorganisms such as bacteria, sperm cells, and various kinds of protozoa move in the low $Re$ number regime. In such a regime, linear viscous forces dominate over nonlinear inertial forces~\cite{childress1981,brennen1977,vogel1994} and locomotion must result from non-reciprocal motion in order to break time-reversal symmetry~\cite{qian2008,yu2006}; this is the so-called ``scallop theorem''~\cite{purcell1977}.

More than fifty years ago, G.I. Taylor beautifully demonstrated that a slender body could swim at low $Re$ number by generating traveling waves along its body~\cite{taylor1951}. An attractive example of such a low $Re$ number swimmer is found in the nematode (roundworm) \emph{Caenorhabditis elegans}. This small free-living swimmer ($\approx1$~mm long) is a well established model organism used extensively for genetic research~\cite{brenner1974} due in part to its short life cycle and fully sequenced genome~\cite{consortium1998}. The motility of \emph{C. elegans} results from the coordinated activity of 95 body wall muscle (BWM) cells, which are highly similar in both anatomy and molecular makeup to vertebrate skeletal muscles~\cite{white1976,white1986}. Anatomically, such BWM cells are arranged longitudinally in four quadrants along the nematode body,  with two rows of muscle cells (dorsal and ventral) in each quadrant~\cite{liu2006,boyle2008}. Experimental observations have shown that motility of swimming \emph{C. elegans} results from the propagation of bending waves along the nematode's body length~\cite{korta2007,pierce2008}, which is an example of non-reciprocal motion. These bending waves consist of alternating phases of dorsal and ventral muscle contractions driven by the neuromuscular activity of BWM cells~\cite{white1986}. The adaptability observed in the locomotory gait in \emph{C. elegans} has been hypothesized to be encoded in sensory and motor systems (mechanosensation) which allow the nematode to respond to its own movement in different physical surroundings~\cite{korta2007}. In turn, \emph{C. elegans} is known to switch between distinct forms of locomotion when transitioning between locomotion on solid substrates (e.g. crawling) and swimming in liquid environments~\cite{pierce2008}.

The motility behavior of nematodes is a strong function of its material properties such as body stiffness and muscle tonus~\cite{gaugler2004}. While it is generally accepted that during locomotion the nematode's body tissues obey a viscoelastic reaction~\cite{karbow2006,guo2008,fung1993}, quantitative data on \emph{C. elegans'} material properties such as tissue stiffness and viscosity remain largely unexplored. Experimental investigations on locomotive behavior in \emph{C. elegans} have mainly focused on quantifying kinematic properties using metrics such as bending frequency, amplitude, wavelength and nematode's centroid velocity~\cite{korta2007,pierce2008,cronin2005,feng2004,ramot2008}. Only recently have the nematode's material properties begun to be probed. For example, static indentation measurements on anesthetized nematodes using piezoresistive cantilevers~\cite{park2007} have provided experimental values for the effective Young's modulus ($E$) of the \emph{C. elegans'} ``shell'', which is composed of the cuticle, the hypodermis, and longitudinal muscles. Values are found to be on the order of 400 MPa and are much closer to stiff rubber than to soft tissues. More recently, estimates of both Young's modulus and tissue viscosity of \emph{C. elegans} have been obtained using kinematic experimental data combined with a dynamic model of locomotion~\cite{sznitman2010}. The authors find that the values of the nematode's tissue properties ($E\approx4$~kPa) are much closer to biological soft-tissues~\cite{engler2006,yamada2000}. Despite recent efforts, however, there is still a dearth of knowledge on overall mechanical properties of \emph{C. elegans}' body and their dynamic response to the external mechanical load imposed during locomotion.

In this paper, we investigate the motility of \emph{C. elegans} as a function of fluid viscosity (i.e. mechanical loading) in both experiments and in a model. In addition, the nematode's {\it effective} material properties including Young's modulus and tissue viscosity are estimated as a function of mechanical loading. For the range of viscosities investigated here (1.0-12~mPa$\cdot$s), experiments show that \emph{C. elegans} swim in a periodic fashion while generating traveling waves of body curvature with amplitudes that decay from head to tail. While the spatial form of the nematode's swimming gait (body amplitude and wavelength) is nearly insensitive to increasing fluid viscosity, its beating frequency decreases to a quasi steady-state value ($\sim 1.7$~Hz) while maintaining a nearly constant swimming speed ($\sim 0.35$~mm/s). A model based on force and moment (torque) balance is able to capture the main features of the experiments. The model is used to estimate both the Young's modulus ($E$) and tissue viscosity ($\eta$) of \emph{C. elegans}. We find that both $E$ and $\eta$ increase with increasing applied mechanical loading ($\mu$) suggesting that nematodes respond to their external environment through dynamic changes in their tissue properties (i.e. muscle tonus).

\section{Experimental Methods}

The experimental configuration (Fig.~\ref{force}a) consists of an acrylic chamber that is $1.5$~mm wide and $1$~mm deep; the chamber is sealed with a thin ($0.13$~mm) cover glass. Shallow chambers are used to minimize three-dimensional motion of the nematode. The chamber is filled with an osmotically balanced aqueous solution of M9 buffer~\cite{brenner1974}, which contains approximately 5 to 10 nematodes per loading. The motion of \emph{C. elegans} is imaged using an inverted microscope and a high-speed camera at 125 frames per second. The high acquisition rate ensures the ability to resolve small linear displacements along the nematode's body between consecutive frames. Wild-type (N2) strains of \emph{C. elegans} are grown and maintained on agar plates using standard culture methods~\cite{brenner1974}. They are fed with the \emph{E. coli} strain OP50. All experiments are performed on hypochlorite synchronized young adult animals. Strains are obtained from the Caenorhabditis elegans Genetic Stock Center.

Figure~\ref{fig_1}(b) shows a snapshot of a typical wild-type \emph{C. elegans} swimming in a buffer solution. Nematode kinematics are characterized using in-house image analysis codes. They are used i) to create continuous two-dimensional (2D) skeletons from segmented body shapes and ii) to calculate the nematode's center of mass (centroid) position. The nematode is tracked over multiple bending cycles, and the skeletons represent the nematode's body centerlines. Results show that the nematode moves in a highly periodic fashion, as indicated by the sweeping motion of its tail (Fig.~\ref{fig_1}b). In particular, body displacement amplitudes of the head sweeping motion are always larger than tail amplitudes. In the buffer solution ($\mu \approx 1.0$~mPa$\cdot$s), head and tail amplitudes are typically 490~$\mu$m and 375~$\mu$m, respectively.

The nematode's swimming speed ($U$) is calculated by differentiating the nematode centroid position over time. The beating frequency $f=2\pi/\omega$ and period $T=1/f$ are calculated from the measured nematode's curvature $\kappa(s,t)$~\cite{sznitman2010}. Here, $\omega$ is the angular frequency. The curvature is defined as $\kappa(s,t)=d\phi/ds$, where $\phi$ is the angle made by the tangent to the $x$-axis at each point along the nematode's centerline, and $s$ is the arc-length coordinate spanning the nematode's head ($s=0$) to its tail ($s=L$).

Fluids of different viscosities $\mu$ are prepared by adding small amounts of high molecular weight (MW) polymer to the M9 buffer solution. The polymer of choice is carboxy-methyl cellulose (CMC, MW=2.1 x 10$^{6}$). Fluid viscosity is varied by changing the polymer concentration in the buffer solution from 0 ppm to 2000 ppm. The fluid viscosities range from $1.0~$mPa$\cdot$s to $12~$mPa$\cdot$s corresponding to the buffer and the 2000 ppm solutions, respectively (Fig.~\ref{fig_2}). Since CMC is a flexible polymer, it is natural to expect viscoelastic effects to arise in the fluid~\cite{arratia2005,arratia_b2005}. However, the salt ions in the buffer solution screen the CMC molecules. This screening tends to keep the molecule in the coiled state even at moderate strain rates ($\sim 10~s^{-1}$). As a result, viscoelastic and strain-rate dependent viscosity behaviors are minimized. Figure~\ref{fig_2} shows the fluid shear-viscosity versus shear-rate. Note that the shear viscosities are nearly-constant even for the highest concentration solutions. The shaded area in Fig.~\ref{fig_2} corresponds to the mean shear-rates observed in the swimming experiments. No appreciable first normal stress-difference (N1) is observed for all solutions investigated here.

\section{Kinematics of Undulatory Swimming}

The swimming kinematics of \emph{C. elegans} are investigated as a function of increasing fluid viscosity ($\mu$). Figure~\ref{fig_3} shows (i) the evolution of the nematode's skeletons over a representative bending cycle and (ii) the spatio-temporal evolution of the nematode's body curvature $\kappa(s,t)$ in the buffer (Fig.~\ref{fig_3}a and b) and in the 1000 ppm solution (Fig.~\ref{fig_3}c and d). Envelopes of skeleton position are aligned with respect to their centroid, and show that the spatial form of the nematode's swimming gait is nearly independent of fluid viscosity (Fig.~\ref{fig_3}a and c). The nematode's wavelength is nearly constant as well, with a value of approximately one body length for the viscosity range investigated here.

Figure~\ref{fig_3}(b) and (d) shows approximately nine (9) bending cycles of body curvature $\kappa$ for the buffer solution and 1000 ppm cases, respectively. The curvature values are color-coded such that red and blue represent positive and negative values of $\kappa$ in mm$^{-1}$, respectively. The $y$-axis corresponds to the non-dimensional position $s/L$ along the length of the nematode's body. The contour plots show the existence of periodic, well-defined diagonally oriented lines. These diagonal lines are characteristic of bending waves, which propagate in time along the nematode body length. The waves of body curvature travel along the nematode and the curvature magnitude decays from head to tail. This behavior is observed both in the buffer solution (Fig.~\ref{fig_3}b) and in the 1000 ppm case (Fig.~\ref{fig_3}d).

Next, the nematode's swimming behavior is characterized as a function of fluid viscosity or mechanical load. Figure~\ref{fig_4} shows the average swimming speed $U$ (Fig.~\ref{fig_4}a), the beating frequency $f$ (Fig.~\ref{fig_4}b), and the body displacement amplitudes $A$ (Fig.~\ref{fig_4}c) as a function of fluid viscosity. A population of up to twenty (n=20) individual nematodes is used to experimentally measure the above quantities for each fluid viscosity ($\mu$). Surprisingly, we find that the nematode's forward speed is nearly independent fluid viscosity. For all fluids, the nematode's average speed is approximately $0.35 \pm 0.01~$mm/s. Such speed corresponds to a Reynolds number ranging from 0.03 to 0.4; for the range of fluid viscosities investigated here, the $Re$ is below unity. Thus, the motility of \emph{C. elegans} can be considered to be viscous dominated even in the buffer solution.

By contrast, the nematode's beating frequency ($f$) decreases with increasing fluid viscosity ($\mu$), as shown in Fig.~\ref{fig_4}(b). The values of $f$ seem to reach an asymptotic value of approximately 1.66~Hz as $\mu$ is increased to a maximum value of $12~$mPa$\cdot$s. The absolute change in frequency, however, is relatively small such that a ten-fold increase in $\mu$ yields an effective reduction of approximately 20$\%$ in $f$, from $2.02 \pm 0.04$~Hz to $ 1.66 \pm 0.04~$Hz. The spatial form of the nematode's swimming gait seems to be unaffected over the range of viscosities; body displacement amplitudes remain near 0.25~mm (Fig.~\ref{fig_4}c). Such findings suggest that \emph{C. elegans} maintains a consistent swimming gait as it senses its fluidic environment~\cite{pierce2008}, at least for the range of fluid viscosities investigated here. The data in Fig.~\ref{fig_4} indicates that \emph{C. elegans} may respond to the increased mechanical load by adapting its swimming kinematics to maintain a constant swimming speed. This adaptability suggests that \emph{C. elegans} may not be power-limited~\cite{korta2007}.


\section{\label{sec:level1} Swimming Model}
In this section, a simple model is developed to capture the dynamics of the swimming nematode. This model will be later coupled with the kinematic data in order to estimate the nematode's material properties including its Young's modulus $E$ and its effective tissue viscosity $\eta$ as a function of fluid viscosity. We begin by considering the motion of the nematode in the limit of low $Re$ numbers~\cite{grayhancock1955}. The ratio of its body length ($L$) to diameter ($D$) is $L/D \approx 15$ such that the nematode can be approximated as a slender body. The nematode's motion is described in terms of its center-line ${y}(s,t)$, where $s$ is the arc-length along the nematode's body and $t$ is time. The swimming nematode experiences no net total force or torque (moments) such that, in the absence of inertia, the equations of motion become
\begin{eqnarray}
 \frac{\partial \vec{F}}{\partial s} & =& C_t \vec{u}_t
 + C_n \vec{u}_n, \label{force} \\
 \frac{\partial M}{\partial s}& =&-[F_{y}\cos(\phi)-F_{x}\sin(\phi)] \label{moment}.
 \end{eqnarray}

In Eq.~(\ref{force}), $\vec{F}(s,t)$ is the internal force in the nematode, $C_{i}$ is the drag coefficient experienced by the nematode, $\vec{u_{i}}$ is the nematode velocity, and the subscripts $t$ and $n$ correspond to the tangential and normal directions, respectively. The drag coefficients $C_{t}$ and $C_{n}$ are obtained from resistive force theory~\cite{grayhancock1955}. Due to the finite confinement of nematodes between parallel walls, corrections for wall effects on the resistive coefficients are estimated for slender cylinders \cite{brennen1977,katz1975}.

In Eq.~(\ref{moment}), $M=M_{p}+M_{a}$, where $M_{p}$ is a passive moment and $M_{a}$ is an active moment generated by the muscles of the nematode; the active and passive moments are parts of a total internal moment~\cite{karbow2006,guo2008}. The passive moment is described using the viscoelastic Voigt model \cite{fung1993} such that $M_{p} = EI\kappa + \eta_{p} I(\partial\kappa/\partial t)$, where $I$ is the second moment of inertia of the nematode cross section and $\kappa(s,t)$ is the curvature along the nematode. Qualitatively, the elastic part of the Voigt model is represented by a spring of stiffness $E$ while the dissipative part of the Voigt model is represented by a dashpot filled with a fluid of viscosity $\eta_p$. More detail on the model may be found elsewhere~\cite{sznitman2010}.

In the present model, it is assumed that two homogeneous {\it effective} material properties describe the entire nematode body, namely (i) a constant Young's modulus $E$ and (ii) a constant tissue viscosity $\eta$. In reality, the anatomy of the nematode is more complex; its body plan consists of an outer tube (or ``shell'') separated from an inner tube by a fluid-filled pseudocoelom~\cite{park2007}. In particular, \emph{C. elegans} has a shell-type hydrostatic skeleton~\cite{harris1957}. The outer tube is comprised of various organs including the cuticle, hypodermis, excretory system, neurons and BWM cells, while the pharynx, intestine, and gonad form the inner tube which remains under pressure. For simplicity, however, the complex internal structure of \emph{C. elegans} is treated as lumped into two single material properties ($E$ and $\eta$).  Also, the nematodes have tapered body shapes at their head and tail, such that the bending modulus ($K_B=I\sqrt{E^{2} + \omega^{2}\eta^{2}}$) varies along the body. In the present model, however, it is assumed that the nematode's cross-section is constant such that values of $K_B$ remain constant as a function of the nematode's arc length $s$.

The active moment generated by the muscle is defined as $M_{a} = -(EI\kappa_{a} + \eta_{a}I \partial\kappa/\partial t)$, where $\kappa_{a}$ is a position and time dependent preferred curvature produced by the muscles of the nematode and $\eta_{a}$ is a positive constant~\cite{thomas1998}. A simple form for $\kappa_{a}$ can be obtained by assuming that $\kappa_{a}$ is a sinusoidal function of time with an amplitude that decreases from the nematode's head to its tail~\cite{sznitman2010}. Note that if $\eta = \eta_{p} - \eta_{a} > 0$, there is net dissipation of energy in the tissue; conversely, if $\eta = \eta_{p} - \eta_{a} < 0$, there is net generation of energy in the tissue. For live nematodes, it is expected that $\eta = \eta_{p} - \eta_{a} < 0$ because the net energy produced in the (muscle) tissue is needed to overcome the drag from the surrounding fluid.

Equations~(\ref{force}) and (\ref{moment}) are simplified by noting that the nematode moves primarily in the $x$-direction (see Fig.~\ref{fig_1}b). While {\it C. elegans} may generate relatively large body displacement amplitudes ($A/L\sim0.25$), deflections of its centerline from the $x$-axis are assumed to be small such that the influence of non-linearities is neglected. The small deflection or displacement assumption is necessary in order to obtain a closed-form solution for the curvature $\kappa(s, t)$. Assuming small deflections from the centerline, $s \approx x$ and $\cos(\phi)\approx 1$. This results in a linearized set of equations given by
\begin{eqnarray}
\frac{\partial F_{y}}{\partial x}-C_{n}\frac{\partial y}{\partial t}=0,
\label{eq:yforbal} \\
\frac{\partial M}{\partial x}+F_{y}=0. \label{eq:mombal}
\end{eqnarray}
Differentiating Eq.~(\ref{eq:mombal}) with respect to $x$ and combining with Eq.~(\ref{eq:yforbal}), we obtain
\begin{equation}
\frac{\partial^{2} M}{\partial x^{2}}+C_{n}\frac{\partial y}{\partial t}=0. \label{eq:final}
\end{equation}

The boundary conditions are such that both the force and moment at the nematode's head and tail are equal to zero. That is, $F_{y}(0,t)=F_{y}(L,t)=0$ and $M(0,t)=M(L,t)=0$. Note that the zero moment boundary conditions at the head and tail imply that $EI\kappa(0,t) + \eta I(\partial\kappa(0,t)/\partial t) = EI\kappa_{a}(0,t)$ and $EI\kappa(L,t) + \eta I(\partial\kappa(L,t)/\partial t) = EI\kappa_{a}(L,t)$.

Experiments show that the curvature $\kappa$ has non-zero amplitudes both at the head ($x=0$) and the tail ($x=L$). In order to capture this observation, we assume that $\kappa_{a}(x,t)$ is a sinusoidal wave with decreasing amplitude of the form $\kappa_{a}(x,t) = Q_{0}\cos\omega t + Q_{1}x\cos(\omega t - B)$, where $Q_{0}$, $Q_{1}$ and $B$ are inferred from the experiments~\cite{sznitman2010}. If the curvature amplitude at the head is larger than that at the tail, then the nematode swims forward. Conversely, if the curvature amplitude is smaller at the head than at the tail, the nematode swims backward; if the amplitudes are equal at the head and tail then it remains stationary.

Equation~(\ref{eq:final}) is solved for $y(x,t)$ in order to calculate the nematode's curvature, which is defined as $\kappa(x,t)= \partial^{2}y/\partial x^{2}$. The solution for $y(x,t)$ is a superposition of four traveling waves of the general form $A_{i}\exp(-\beta x\cos P_{i})\cos(\beta x\sin P_{i}-\omega t - \phi_{i})$ where $\beta$ and $P_{i}$ can be written in terms of $\omega$, $E$, and $\eta$, while $A_{i}$ and $\phi_{i}$ are constants to be determined by enforcing the boundary conditions discussed above (for more details, please see~\cite{sznitman2010}). Note that the formulation presented here does not assume a wave functional form for $\kappa(x,t)$. Rather, the wave is obtained as part of the solution.

The nematode body position $y(x,t)$ predicted by the model is fitted to the experimental data in order to compute the curvature $\kappa$. In short, time profiles (about $8$ seconds long) of the motion are fitted at several points along the nematode body using a Levenberg-Marquardt algorithm. The fitting method yields estimates of the bending modulus ($K_{b}$) and the phase angle $\psi = \tan^{-1}(\eta\omega/E)$. The estimated values of $K_{b}$ and $\psi$ are later used to calculate the nematode's Young's modulus and tissue viscosity. Curvature $\kappa$ contour plots for the buffer solution case obtained from experiments and from the model are shown in Fig.~\ref{fig_5}(a) and Fig.~\ref{fig_5}(b), respectively. For the particular nematode shown in Fig. 5, the estimated values of $K_{b}$ and $\psi$ are $1.87 \times 10^{-15}~$Nm$^2$ and $-78.9^o$, respectively. The overall estimated values of the bending modulus $K_b$ and phase angle $\psi$, for the buffer solution case, using up to 20 nematodes is $K_b = 2.61 \times 10^{-15} \pm 0.22 \times 10^{-15}$~Nm$^2$ and $\psi=-73.4^o \pm 1.7^o$. As shown in Fig. 5(a) and 5(b), the model is able to capture the main qualitative features of the curvature profile including the decay from head to tail.

Next, we compare the temporal evolution of curvature $\kappa(t)$ obtained from the experiments and the model as a function of fluid viscosity $\mu$ at a single body location $s/L = 0.5$. Values of $\kappa(t)$ for the buffer solution, 1000 ppm, and 2000 ppm cases are shown in Fig.~\ref{fig_5}(c), 5(d), and 5(e), respectively. The model seems to be able to capture the experimentally measured values of the frequency and amplitude of $\kappa(t)$. The relative error between the values of $\kappa(t)$ obtained from the experiments and the model is computed as $\epsilon=1-[(\sum_{i}|\kappa_{model}(t_{i})|)/\sum_{i}|\kappa_{exp}(t_{i})|]$. The values of $\epsilon$ are $3.27\%$, $1.96\%$, and $2.31\%$ for the buffer, 1000 ppm and 2000 ppm cases, respectively. These values indicate that the analytical solution presented here, while not perfect, is able to capture the main features observed in experiments.

\section{\emph{C. Elegans}' Material Properties}
Next, the \emph{effective} material properties of nematodes ($E$ and $\eta$) are estimated from the definition of the bending modulus $K_b$ and phase angle $\psi$. Here, the nematode is modeled as an idealized cylindrical geometry~\cite{park2007,zelenskaya2005}, where the effects of internal hydrostatic pressure are neglected; this pressure contribution has been recently shown to have only modest effects on nematode body stiffness~\cite{park2007}. In particular, we consider the second moment of inertia $I$ of the nematode cross section as represented by the \emph{C. elegans}' ``shell'' which comprises of the cuticle, the hypodermis, and longitudinal muscles. This shell-like structure is defined as $I = \pi(r_{o}^4-r_{i}^{4})/4$, where the outer radius of the shell corresponds to the mean nematode radius $r_{o} \approx 35~\mu$m and the inner radius of the shell is assumed to be $r_{i}\approx r_{o}/2$ from electron micrographs of the nematode cross-section~\cite{park2007}.

Results show that, for the nematodes swimming in buffer solution, $E=0.62 \pm 0.05$~kPa and $\eta=-177.1 \pm 15.2$~Pa$\cdot$s. The estimated value of $E$ lie within the wide range of values of tissue elasticity measured for soft tissues, including isolated brain cells ($0.1-1~$kPa) and muscle cells ($8-17~$kPa)~\cite{engler2006}. The values of $\eta$ for live \emph{C. elegans} are negative because the organism's tissues are
generating rather than dissipating energy~\cite{thomas1998,sznitman2010}. We note, however, that the absolute values of tissue viscosity $|\eta|$ are within the range ($10^2-10^4~$Pa$\cdot$s) measured for living cells~\cite{yamada2000,thoumine1997}.

The values of $E$ and $\eta$ for the nematode swimming in fluids of different viscosities $\mu$ are shown on a log-log phase diagram in Fig.~\ref{fig_6}. The estimated magnitudes of both $E$ and $\eta$ increase monotonically over an order of magnitude with increasing fluid viscosity or mechanical load; each data point corresponds to at least twenty experiments. The data shows an increase in the nematode's muscle tonus as the nematode senses its environment and responds with dynamic changes in its material properties. The increase in the values of the Young's modulus is associated with the shortening of sarcomeres and higher density of muscle cells typical in muscle contraction~\cite{tawada1990}. The increase in the absolute values of tissue viscosity $|\eta|$ is associated with the larger amount of energy necessary to overcome the increased fluid drag or the energy dissipation due to higher fluid viscosity $\mu$.

\section{Conclusions}

In summary, we have characterized the effects of mechanical loading on the swimming behavior of \emph{C. elegans} at low $Re$ number by varying the fluid viscosity ($\mu$). Results show that over the range of viscosities tested, nematodes swim consistently with a periodic swimming behavior illustrating travelings waves that decay from the nematode's head to tail. The spatial form of \emph{C. elegans}' swimming gait is nearly unchanged while its beating frequency decreases only slightly with increasing viscosity. Given such kinematics, nematodes are nevertheless able to maintain constant their average forward swimming speed. Such findings support altogether \emph{C. elegans}' locomotive adaptability to its surrounding environment.

By coupling experiments with a linearized model based on force and torque balances, we are able to estimate, \emph{non-invasively}, the nematode's tissue material properties such as Young's modulus ($E$) and viscosity ($\eta$). We find that \emph{C. elegans} behaves effectively as a viscoelastic material with $E=0.62 \pm 0.05$~kPa and $\eta=-177.1 \pm 15.2$~Pa$\cdot$s for the buffer solution case. Both the magnitude of the Young's modulus and the tissue viscosity increase over an order of magnitude as fluid viscosity is increased; values remain within the range of that measured for living cells. We speculate that the increase in Young's modulus may be associated with the shortening of sarcomeres and higher density of muscle cells typical in muscle contraction; usually muscle stiffens as larger mechanical loading is applied. In addition, we suspect that the increase in the absolute value of tissue viscosity $|\eta|$ is associated with the larger amount of energy necessary to overcome the increased fluid drag or the energy dissipation due to larger fluid viscosity $\mu$.

\subsection*{Acknowledgements}
The authors would like to thank T. Lamitina for helpful discussions. Some nematode strains used in this work were provided by the Caenorhabditis Genetics Center, which is funded by the NIH National Center for Research Resources (NCRR).

\bibliographystyle{plain}

\newpage
\section*{Figure Legends}
\begin{figure*}[p]
\begin{center}
\includegraphics[width=1.0\textwidth]{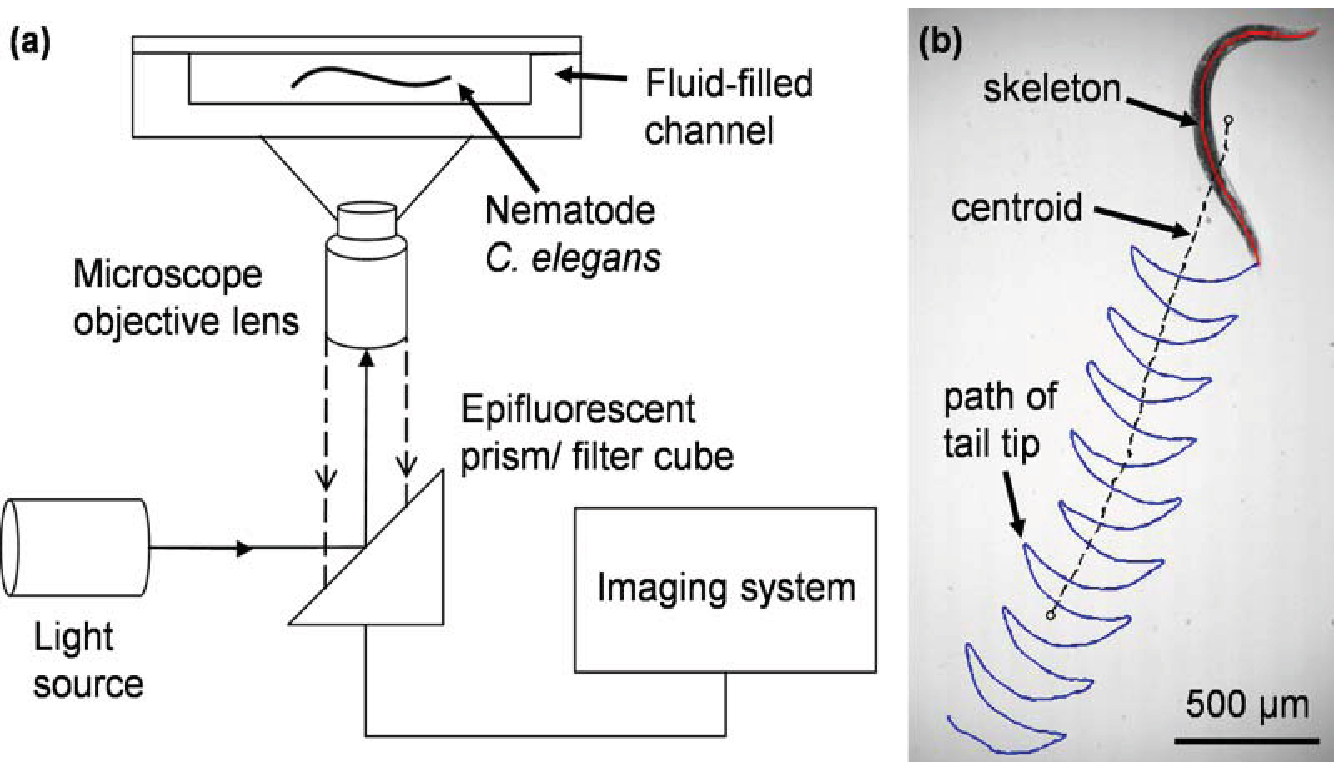}
\caption{}
\label{fig_1}
\end{center}
\end{figure*}

~\\\textbf{Figure~\ref{fig_1}:}\\
\textbf{(a)} Schematic diagram of the experimental setup. The nematode \emph{Caenorhabditis elegans} (\emph{C. elegans}) is placed in a shallow channel that is 1.5~mm wide and 500~$\mu$m deep containing an aqueous solution. The channel is sealed with a thin (0.13~mm) cover glass. \emph{C. elegans} is imaged using standard microscopy and a high-speed camera. \textbf{(b)} Visualization of instantaneous body centerline or nematode's skeleton with resulting centroid and tail-tip trajectories over multiple body bending cycles. Image acquisition rate is 125 frames per seconds. The nematode's undulations lie within the microscope focal plane.

\begin{figure*}[p]
\begin{center}
\includegraphics[width=0.9\textwidth]{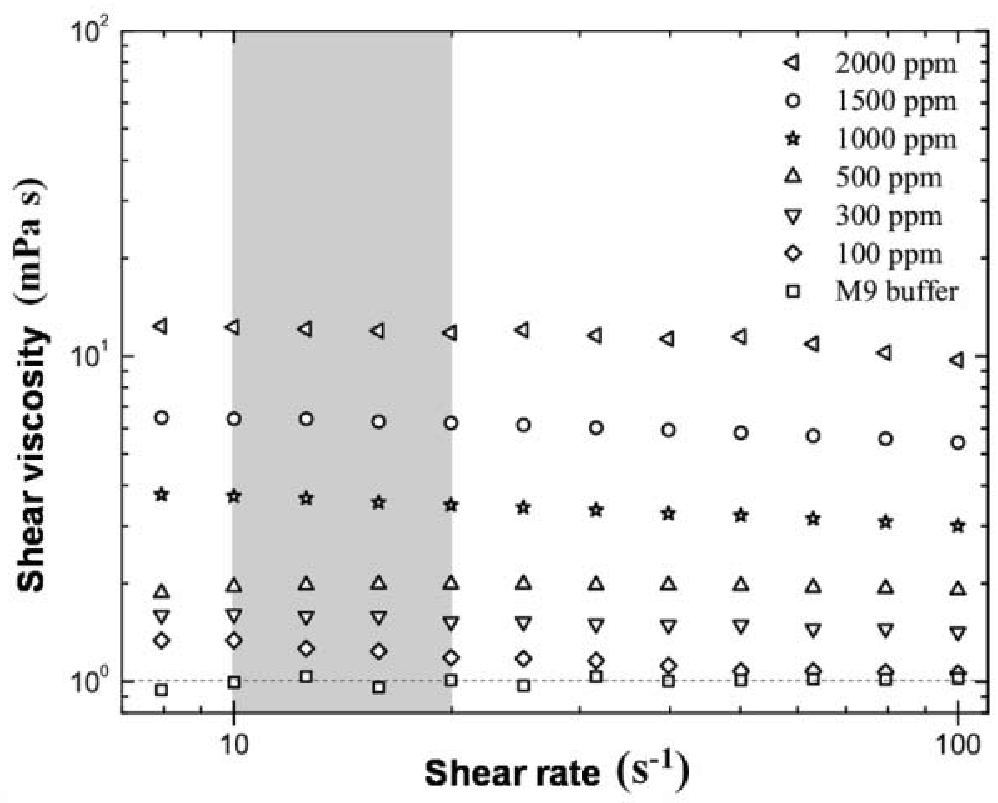}
\caption{}
\label{fig_2}
\end{center}
\end{figure*}

~\\\textbf{Figure~\ref{fig_2}:}\\
Fluid characterization using a stress-controlled rheometer. Shear-viscosity $\mu$ as a function of shear-rate for both buffer and polymeric solutions. Shaded area in the plot corresponds to shear rates ($\dot{\gamma}$) associated with \emph{C. elegans} swimming motion.  For the range of shear rates considered, the values of fluid viscosity are nearly constant and fluid solutions may be regarded as effectively Newtonian. No appreciable normal first normal stress difference (N1) was observed even for the highest polymer concentration (2000 ppm).

\begin{figure*}[p]
\begin{center}
\includegraphics[width=0.95\textwidth]{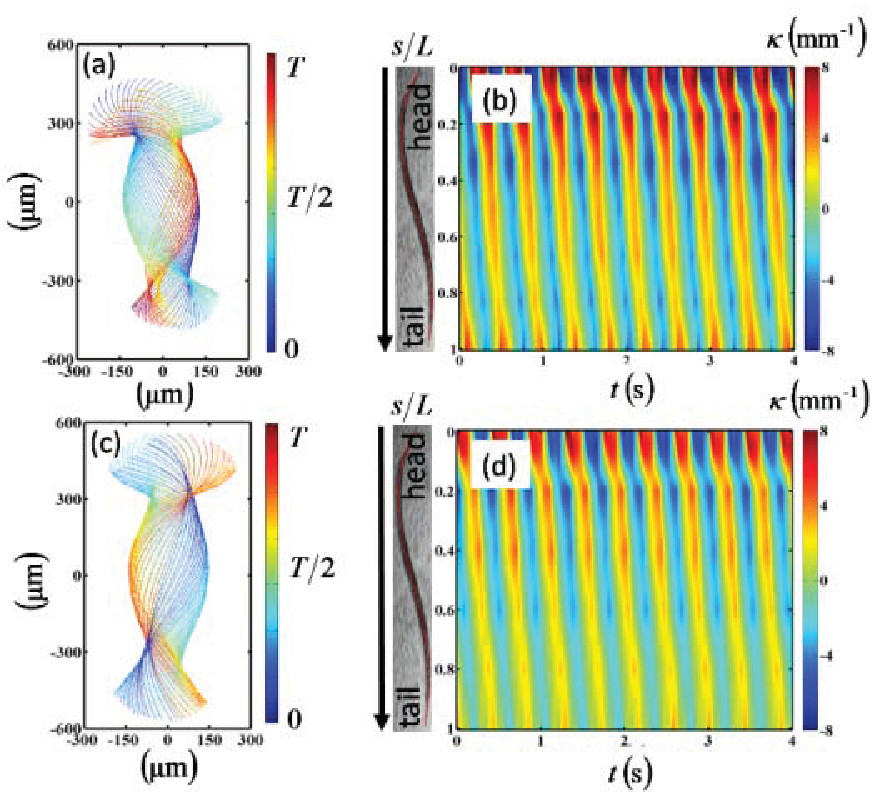}
\caption{}
\label{fig_3}
\end{center}
\end{figure*}

~\\\textbf{Figure~\ref{fig_3}:}\\
Skeletons and spatio-temporal evolution of \emph{C. elegans}' curvature swimming in different fluid viscosities. Examples of the evolution of nematode skeletons over one representative bending cycle in (\textbf{a}) the buffer M9 solution ($\mu \approx 0.8$~mPa$\cdot$s) with $T \approx 0.49~$s, and (\textbf{c}) in the 1000 ppm CMC solution ($\mu \approx 3.59$~mPa$\cdot$s) with $T \approx 0.62~$s. The data reveals a well-defined envelope of elongated body shapes with a wavelength corresponding approximately to the nematode's body length. Envelope shapes are aligned by their centroids and are nearly invariant with increasing viscosity. (\textbf{b}) and (\textbf{d}) Representative contour plots of the \emph{experimentally measured} curvature ($\kappa(s,t)$) along the nematode's body centerline for (\textbf{a}) the buffer solution and (\textbf{d}) 1000 ppm CMC solution, respectively. Approximately $9$ bending cycles are shown over 4 seconds. Red and blue colors correspond to positive and negative $\kappa$  values, respectively. The units of $\kappa$ are in mm$^{-1}$. The $y$-axis corresponds to the dimensionless position $s/L$ along the \emph{C. elegans}' body length where $s=0$ is the head and $s=1$ is the tail.

\begin{figure*}[p]
\begin{center}
\includegraphics[width=0.85\textwidth]{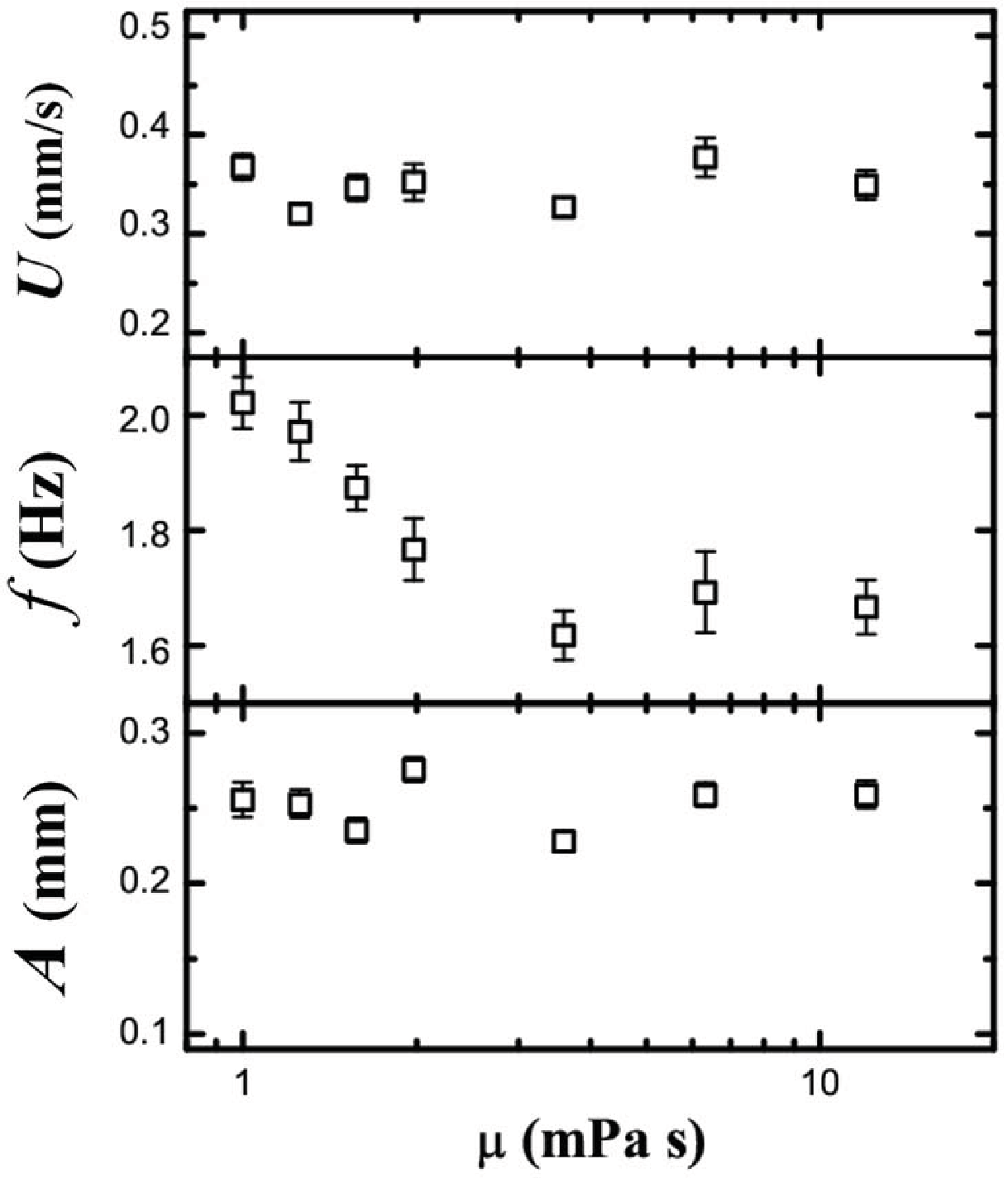}
\caption{}
\label{fig_4}
\end{center}
\end{figure*}

~\\\textbf{Figure~\ref{fig_4}:}\\
Nematode tracking measurements as a function fluid viscosity $\mu$. Nematode's (\textbf{a}) swimming speed ($U$), (\textbf{b}) beating frequency ($f$), and (\textbf{c}) body displacement amplitude ($A$). Both the forward speed and displacement amplitudes are nearly constant with increasing fluid viscosity. The nematode's bending frequency decreases to a quasi-steady value with increasing fluid viscosity or mechanical load.

\begin{figure*} [p]
\begin{center}
\includegraphics[width=1.0\textwidth]{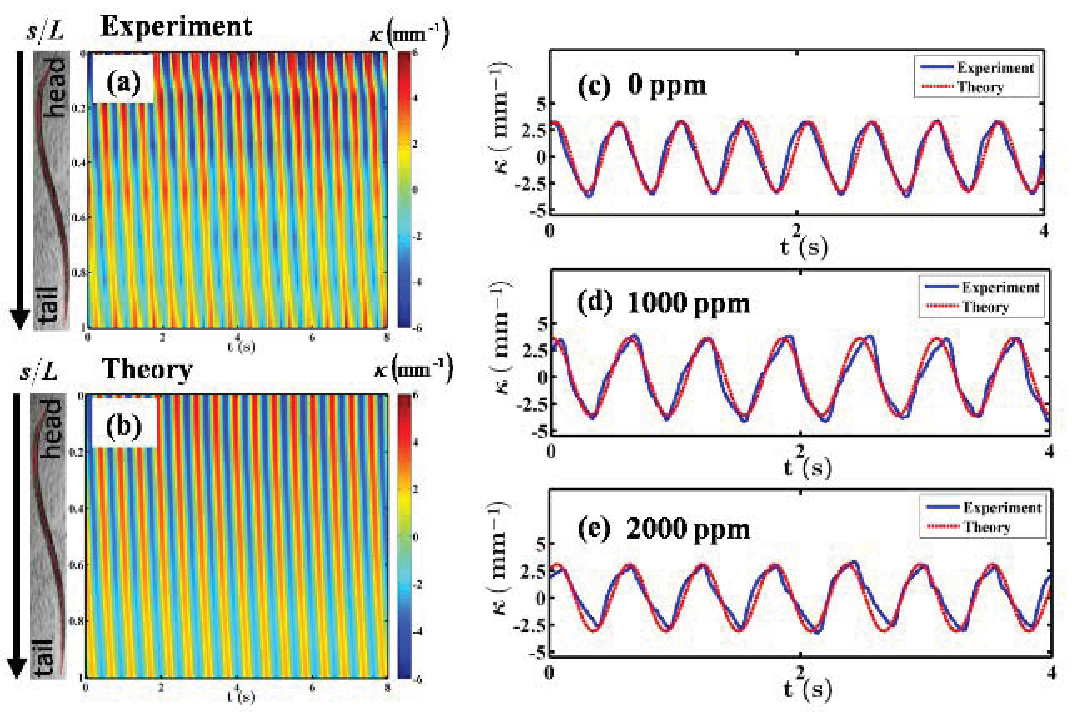}
\caption{}
\label{fig_5}
\end{center}
\end{figure*}

~\\\textbf{Figure~\ref{fig_5}:}\\
(\textbf{a}) Representative contour plot of experimentally measured curvature $\kappa (s,t)$ along the nematode's body centerline for the buffer solution case. \textbf{(b)} Corresponding contour plot of curvature values obtained from the model. The model captures the longitudinal bending wave with decaying magnitude, which travels from head to tail. For the nematode shown, estimates of the bending modulus and phase angle yield $K_b = 1.87 \times 10^{-15}~$Nm$^2$ and $\psi = -78.9^o$, respectively. Such values correspond to $E=0.32~$kPa and $|\eta| = 134.5~$Pa$\cdot$s. (\textbf{c})-(\textbf{e}) Comparison between experimental and theoretical curves of $\kappa (s,t)$ at $s/L=0.5$ for the buffer solution, 1000 ppm, and 2000 ppm cases, respectively. The relative error between the model and the experiments is less than $4\%$ (see text).

\begin{figure*} [p]
\begin{center}
\includegraphics[width=1.0\textwidth]{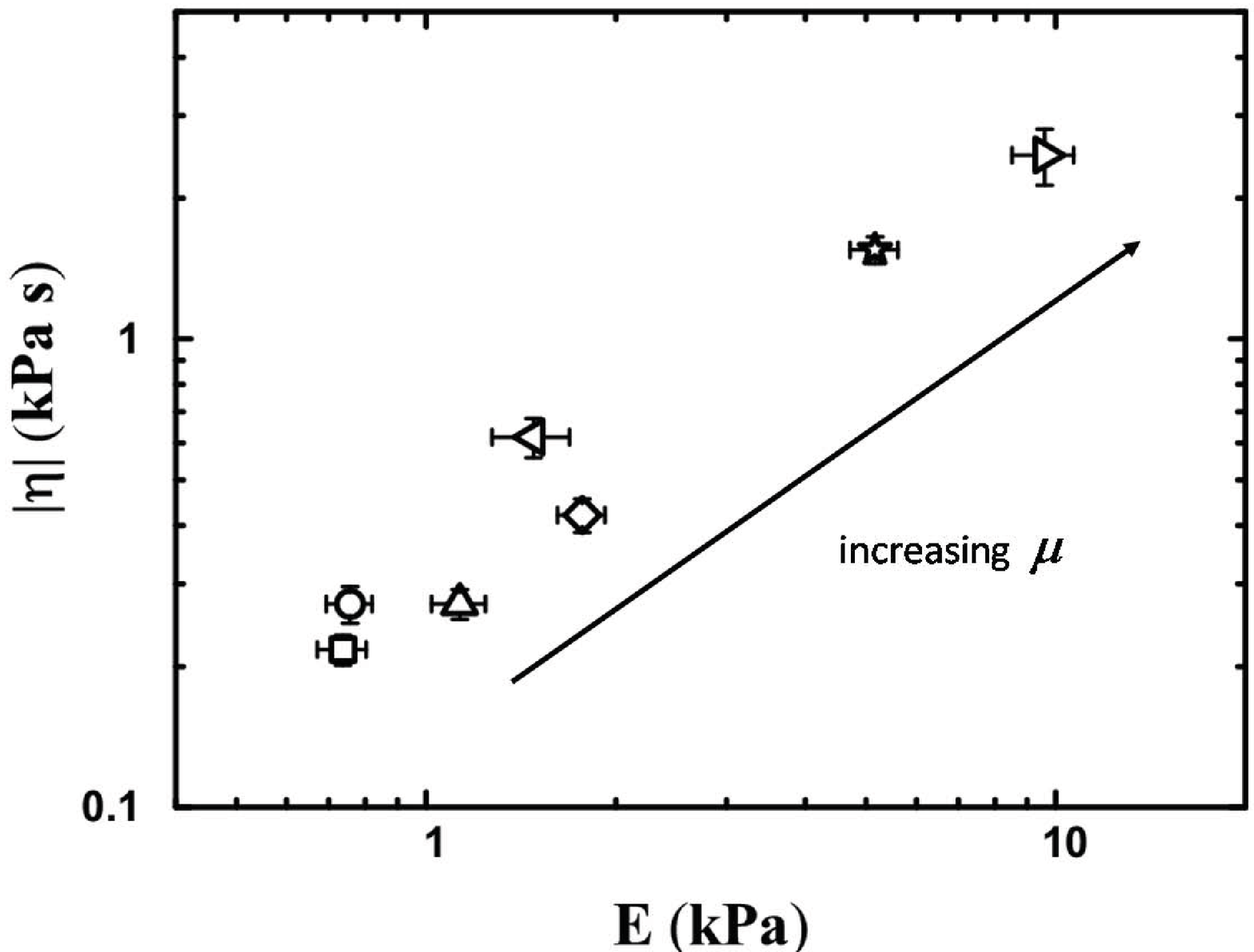}
\caption{}
\label{fig_6}
\end{center}
\end{figure*}

~\\\textbf{Figure~\ref{fig_6}:}\\
\emph{C. elegans}' estimated material properties as a function of fluid viscosity. The estimated Young's modulus $E$ and absolute values of tissue viscosity $|\eta|$ increase monotonically with increasing fluid viscosity.

\newpage

\newpage

\newpage

\newpage

\newpage

\newpage

\end{document}